\documentclass[times]{speauth}

\usepackage{moreverb}
\usepackage{listings}
\usepackage{amsmath}
\usepackage{cases}
\usepackage{multirow}
\usepackage{blkarray}
\usepackage{subfig}
\usepackage{color}
\newcommand\BibTeX{{\rmfamily B\kern-.05em \textsc{i\kern-.025em b}\kern-.08em
T\kern-.1667em\lower.7ex\hbox{E}\kern-.125emX}}

\def\volumeyear{2016}

\usepackage{graphicx}

\begin{document}

\runningheads{L.~H.~Nunes}{A demonstration of the \journalabb\
class file}

%\title{Multi-Objective Internet of Things Resource Discovery through Pareto-optimality Criterion}

\title{Multi-criteria IoT Resource Discovery: A Comparative Analysis}

\author{Luiz H. Nunes\affil{1,2}, Julio C. Estrella \affil{1}, Charith Perera\affil{3}, Stephan Reiff-Marganiec\affil{4} and Alexandre N. Delbem\affil{1}}

\address{\affilnum{1} University of S\~ao Paulo (USP), Institute of Mathematics and Computer Science (ICMC), S\~ao Carlos-SP, Brazil \break
\affilnum{2} Instituto Federal de S\~ao Paulo (IFSP), Mat\~ao-SP, Brazil \break
\affilnum{3}Faculty of Maths, Computing and Technology, The Open University, Walton Hall, Milton Keynes, MK7 6AA - UK \break
\affilnum{4} University of Leicester, University Road, Leicester, LE1 7RH - UK}

\corraddr{ University of S\~ao Paulo (USP), Institute of Mathematics and Computer Science (ICMC), S\~ao Carlos-SP, Brazil. E-mail: lhnunes@icmc.usp.br}

%1. Clearly state your contribution in this work. I need to be sure about that before making any decision.
%2. Clearly state the experimental setup. I can see some information in Tables 1-2, but I still need much better explanation about the "case" and what exactly are you looking for. How quality is expressed in each case? What a "good solution" looks like for your case study.
%3. How the Pareto is found and calculated?
%4. Figures are not clear enough to understand what these lines actually mean. - OK
%5. Discuss the results.
%EXTRA - Rewrite part of results section

\begin{abstract}

The growth of real world objects with embedded and globally networked sensors allows to consolidate the Internet of Things paradigm and increase the number of applications in the domains of ubiquitous and context-aware computing. The merging between Cloud Computing and Internet of Things named Cloud of Things will be the key to handle thousands of sensors and their data. One of the main challenges in the Cloud of Things is context-aware sensor search and selection. Typically, sensors require to be searched using two or more conflicting context properties. Most of the existing work uses some kind of  multi-criteria decision analysis to perform the sensor search and selection, but does not show any concern for the quality of the selection presented by these methods. In this paper, we analyse the behaviour of the SAW, TOPSIS and VIKOR multi-objective decision methods and their quality of selection comparing them with the \textit{Pareto}-optimality solutions. The gathered results allow to analyse and compare these algorithms regarding their behaviour, the number of optimal solutions and redundancy.

\end{abstract}


\maketitle

%\footnotetext[2]{Please ensure that you use the most up to date class file,
%available from the SPE Home Page at\\
%\href{http://www3.interscience.wiley.com/journal/1752/home}{\texttt{http://www3.interscience.wiley.com/journal/1752/home}}}

\vspace{-6pt}
\vspace{-0.6cm}

\section{Introduction}
\vspace{-2pt}
%The recently increasing of smart objects connected in the Internet allowed to become the Internet of Things (IoT) in a reality in our lives. According to the Gartner Report, there is about 6.4 billion of connected things moving a market around \$235 billion just with  end-users services. The integration of Cloud Computing and Internet of Things named as Cloud of Things (CoT) is extremely important to handle with the connected things, their data and the provided services.

The Internet of Things (IoT) is an ecosystem that interconnects physical objects with telecommunication networks, joining the real world with the cyberspace and enabling the development of new kinds of services and applications. The IoT world  is composed of small sensors and actuators embedded in the objects such as electronic devices (e.g.~smartphones or tablets), clothes, alarm systems, cars, domestic appliances and industrial machines, which are capable of interacting with each other and with their environment. 

%procurar referencia mais recente que diga a mesma coisa
Recently, the number of devices has grown rapidly and it is anticipated that between 2015 and 2016 about 20 billion devices will be connected to the Internet creating a market of around 91.5 billion dollars \cite{Evans:2011}. These things generate an amount of data which cannot be handled in a standalone power-constrained IoT environment. The integration of IoT with cloud computing, named Cloud of Things (CoT),  can facilitate unprecedented ubiquitous sensing services and powerful resources to process sensing data streams beyond the capability of individual “things” \cite{Aazam:2014}.
%is the key to manage thousands of sensors and their data in a cost-effective manner .

Different domains can benefit from CoT applications such as logistics \cite{Dada:2008, Atzori:2010}, healthcare \cite{Islam:2015}, smart cities \cite{Dada:2008}, environmental monitoring  \cite{Robles:2014, Manna:2014} and assisted driving \cite{Atzori:2010, Gerla:2014}. However, the CoT poses new challenges as it needs to combine different types of services provided by multiple stakeholders and support a large number of users and devices. One of these challenges is to provide a set of tools and environments for development of dynamic applications and ensure their seamless execution to meet the Quality of Context (QoC) and Quality of Service (QoS) requirements imposed by different kinds of applications \cite{Gubbi:2013}.

While the nature of the CoT makes it suitable for provisioning the aforementioned services, ensuring their QoS and QoC requirements imposes complex challenges such as the resource constrained environment, redundant data, heterogeneity of the sensors nodes, dynamic network topology and size, and an unreliable communication medium. These factors can affect the user experience \cite{Bhuyan:2010}. In addition, it is common to find two or more conflicting QoS and QoC requirements in this kind of service.

Thus, several papers such as \cite{Shah:2012,Perera:2013,Gao:2014}  use some kind of Multiple-Criteria Decision Analysis (MCDA) to perform the sensor search and selection to achieve the best trade-off between the QoS and QoC properties. On the other hand, these papers do not show any concern about the quality of the selection presented by these methods regarding aspects like redundancy and dispersion of the selected sensors.

In this paper, we present a qualitative study of three MCDA methods used to establish the relative importance of multiple attributes and alternatives. In particular, we investigate the behaviour of the \textit{Simple Additive Weight method (SAW)}, the \textit{Technique for the Order of Prioritisation by Similarity to Ideal Solution (TOPSIS)} and \textit{VIseKriterijumska Optimizacija I Kompromisno Resenje} (VIKOR) under different conditions. We also analyse the quality of the solutions proposed by these methods comparing them with the available \textit{Pareto} optimal solutions.  The scientific contributions of our work can be summarized as:

\begin{itemize}
\item We proposed a methodology that can be used to compare  different sensor search techniques from a quality of search  perspective. We have demonstrated how our proposed methodology works using three different sensor search techniques.
\item We examined  the impact of  \textit{`number of context-properties'} towards the overall quality of sensor search.
\item We examined the impact of \textit{`number of sensors required to be selected'} towards the overall quality of sensor search.
\item We evaluated and compared the overall quality of sensor search between three different  Multi-Criteria Decision Analysis (MCDA) techniques, namely \textit{Simple Additive Weight method (SAW)}, the \textit{Technique for the Order of Prioritisation by Similarity to Ideal Solution (TOPSIS)} and \textit{VIseKriterijumska Optimizacija I Kompromisno Resenje} (VIKOR).
\end{itemize}

The paper is organized as follows:
Section \ref{sec:MCDM} describes the analysed Multiple-criteria decision-making algorithms 
and the methodology used to evaluate them.
%Section \ref{sec:PerformanceEvaluation} describes the methodology and configurations used for the experiments.
The results are then discussed in Section \ref{sec:Results}. Section \ref{sec:RelatedWork} presents a literature review of existing approaches for sensor search and selection. Finally, the conclusions and directions for future work are presented in Section \ref{sec:Conclusion}.
    
\section{Background}

One of the most accepted definitions of what is IoT is described by Vermesan et al. (2011) \cite{Vermesan:2011}: ``The IoT aims to allow people and things to be connected at any time or place with anything or anyone by any path, network or service''. %IoT encompasses a vast amount of hardware and software technologies, such as Radio-Frequency Identification (RFID), Near Field Communication (NFC), sensors networks, middleware and frameworks \cite{Whitmore:2014}.}
Usually, the IoT is considered as a three-layer architecture, as represented by Figure~\ref{fig:arch}a, showing the perception layer, network layer and application layer \cite{Zhang:Arch:2012, Aazam:2014}. On the other hand, some authors such as Khan et al. (2012) \cite{Khan:2012}, Aazam and Huh (2014) \cite{Aazam:2014} and Fersi (2015) \cite{Fersi:2015} consider two extra layers named as middleware and business, as show in Figure~\ref{fig:arch}b. The main layers objectives can be summarized as:

\begin{figure}[htp]
        \center
        \includegraphics[scale=0.65]{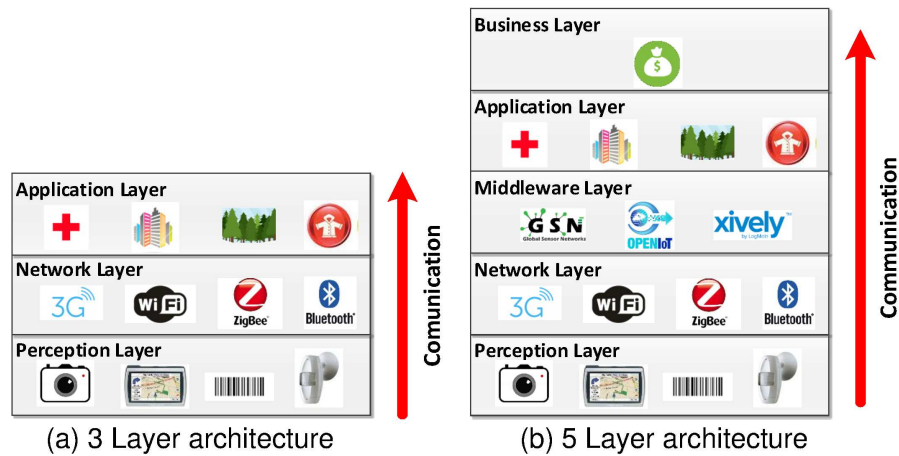}
        \caption{IoT Architectures}
        \label{fig:arch}
\end{figure}

\begin{itemize}
    \item \textbf{Perception layer:} It is main function is to perceive and collect the real environment information and bring them to the virtual environment. Sensors, bar code labels bar, radio-frequency identification devices (RFID), GPS and cameras are concentrated in this layer \cite{Aazam:2014}. These devices can be described by metadata or specific languages such as SensorML, OGC/SWE, SSN W3C, HyperCat and semantic models to enable its use by the up layers \cite{Perera2016}.
    
    \item \textbf{Network layer:} It is responsible for transporting data from the perception layer to be processed. The transmission medium can use wired networks or wireless networks such as 3G, UMTS, Wifi, Bluetooth, infrared and ZigBee depending directly on the types of sensor devices and the environment in which they are deployed \cite{Khan:2012}.
    
    \item \textbf{Middleware layer:} its goal is to offer services and store data received from the network layer. Their services must process the information and make automated decisions based on their results \cite{Khan:2012, Aazam:2014}. Currently, there are several middleware solutions such as GSN \cite{Aberer:2006} and openIoT \cite{Soldatos:2012} to support the management of sensor networks. Usually these solutions are able to abstract the sensors available in the perception layer and offer their resources as a service to end users.
    
    \item \textbf{Application layer:} It presents data from the network layer or middleware layer. This layer must be concerned to present the information according to the specifications or constraints of an user \cite{Aazam:2014}.
    
    \item \textbf{Business layer:}  It is responsible for system management including its applications and services. It defines the business models, graphics and execution flows based on data received from the application layer. The success of IoT depends directly on establishing good business models to analyze the results and determine future business strategies \cite{Khan:2012}.
\end{itemize}

Nowadays, there is much research conducted for the different layers of the IoT architecture, which aim to solve problems related to interoperability, scalability, reliability, data management, privacy and security. One of the most significant challenges involves the middleware layer. Specifically, concentrating on how to support the search and selection of sensors regarding the QoS and QoC properties determined by a user \cite{Gubbi:2013}.

\section{Multiple criteria decision analysis}\label{sec:MCDM}
\vspace{-0.3cm}
Multiple criteria decision analysis (MCDA) refers to making decisions in the presence of multiple, usually conflicting, criteria \cite{Xu:2001}. MCDA algorithms aims to aid in the judgement of the decision making team using a set of objectives and criteria, estimating their relative importance weights and, establishing the contribution of each option regarding to each performance criterion \cite{Dodgson:2009}.

An MCDA problem can be described using an analysis matrix ($M \times N$)  in which element $q_{ij}$ represents the performance of the option according to the decision criterion $c_{j}$ in  different and non-comparable units and scales, as represented in Equation 1. The evaluation matrix is used to represent the relative performance of $q'_{ij}$ using a value/utility function to enable comparisons between the different criteria \cite{Tzeng:2011}.

\begin{align}
    Q = 
    \begin{blockarray}{cccccc} 
    &c_{1} & c_{2} & c_{3} &  & c_{n} \\
        \begin{block}{c[ccccc]}
          q_{1} & q_{11} & q_{12} & q_{13} & \hdots & q_{1n} \\    
          q_{2} & q_{21} & q_{22} & q_{23} & \hdots & q_{2n} \\
                & \vdots &\vdots  &\vdots & \vdots & \vdots \\
          q_{m} & q_{m1} & q_{m2} & q_{m3} & \hdots & q_{mn} \\
        \end{block}
    \end{blockarray} \label{eq1:dm}
 \end{align} 

All MCDA algorithms explicitly define their options and contributions to each criterion, but differs in how they combine the input data. Although MCDA problems are found in different contexts, they usually share common features such as multiple attributes/criteria often forming a hierarchy, conflict among criteria, hybrid nature, uncertainty, large scale and assessments that may not be conclusive \cite{Xu:2001}. Sections \ref{sec:SAW} to \ref{sec:VIKOR} describe three MCDA methods, namely SAW, TOPSIS and VIKOR algorithms for MCDA. 
Section \ref{sec:evaluation} proposes the use of the \textit{Pareto}-Optimality based on the proposed criterion to evaluate the applied MCDAs to select a subset of sensors. 

\subsection{SAW} \label{sec:SAW}

The Simple Additive Weighting (SAW) method is one of the most popular MCDA methods \cite{Memariani:2009, Abdullah:2014}. It provides the additive properties to calculate the final score of alternatives used for weight determinations and preferences, which is the basis of other MCDA methods such as the Analytic Hierarchy Process (AHP) and Preference ranking organization method for enrichment evaluation (PROMETHEE) \cite{Memariani:2009}. According to \cite{Abdullah:2014}, SAW is used in several application domains such as supply chain management, personnel selection problems, project manager selection and facility location selection.

%The Simple Additive Weighting (SAW) 
SAW uses an evaluation score to rank each available option. The score is obtained using a normalized criteria value multiplied by a weight. The options are ranked in descending order according to their final score, which is the sum of the scores for  individual criteria \cite{Afshari:2010}. SAW algorithm can be summarized by the following three steps \cite{Tzeng:2011}:

\begin{enumerate}

\item Normalize the analysis matrix $\textbf{Q}$ described in Equation \ref{eq1:dm} to $\textbf{Q'}$ according to Equation \ref{saw:1a} if the criterion should be maximized or the Equation \ref{saw:1b} if the criterion should be minimized.

\begin{align}
q'_{ij} &=\dfrac{q_{ij} - q_{j}^{min}}{q_{j}^{max} - q_{j}^{min}} &\text{for a criterion to be maximized} \label{saw:1a}\\
q'_{ij} &=\dfrac{q_{j}^{max}- q_{ij}}{q_{j}^{max} - q_{j}^{min}} &\text{for a criterion to be minimized} \label{saw:1b}
\end{align}
        %\begin{equation}\label{saw:1}
        %$q'_{ij} =\dfrac{q_{ij} - q_{j}^{min}}{x_{j}^{max} - q_{j}^{max}}$ , for a criterion to be maximized
        %\end{equation}
        %\begin{equation}\label{saw:2}
        %\\
        %$q'_{ij} =\dfrac{q_{j}^{max}- q_{ij}}{x_{j}^{max} - q_{j}^{max}}$ , for a criterion to be minimized 
        %\end{equation}

\item Compute the score vector $\phi$ of each available option. Each score \textbf{$q'_{i}$} can be calculated using Equation \ref{saw:3}, where w$_{j}$ corresponds to the criterion weight and $N$ represents the number of criteria in the evaluation matrix.

        \begin{align}
            \phi (q_{i}) &= \sum_{j=1}^{N} w_{j} \cdot q'_{ij} \label{saw:3}
        \end{align}

\item Sort options \textbf{$q_{i}$} in decreasing order according to the score $\phi$ ($q'_{i}$) to get the ranking of suitable options.

\end{enumerate}

\subsection{TOPSIS}

TOPSIS explores the attribute information to provide a set of ranked alternatives and requires independent that attribute preferences. The application domains that uses the TOPSIS method has been Supply Chain Management and Logistics, Design, Engineering and Manufacturing Systems,  Business and Marketing Management, Health, Safety and Environment Management, Human Resources Management, Energy Management, Chemical Engineering and Water Resources Management \cite{Behzadian:2012}.
%TOPSIS 
TOPSIS sorts a set of options according to the Euclidean distance from the ideal and negative-ideal solutions. Each option is normalized using a specific criterion value. The ideal solution represents the most desirable level of each criterion across the options under consideration, while the negative-ideal solution reflects the worst-desirable level of each criterion. The options are ranked regarding their closeness to the ideal solution and farness to the negative-ideal solution \cite{Tzeng:2011}. The TOPSIS algorithm can be summarized in the following steps \cite{Opricovic:2004}:

\begin{enumerate}
\item Normalize the analysis matrix $\textbf{Q}$ to $\textbf{Q'}$ according to the Equation \ref{topsis:1}:

    \begin{align} 
        q'_{ij} &=\dfrac{q_{ij}}{\sqrt{\sum_{i=1}^{N}(q_{ij}})^2} \label{topsis:1}
    \end{align}

where N represents the number of options in the evaluation matrix.

\item Determine the  positive ideal points ($p_{+j}$) and the  negative ideal points ($p_{-j}$) of all objective functions using the analysis matrix. For a maximization criterion, the positive ideal and the  negative ideal points can be calculated using Equations \ref{topsis:2a} and \ref{topsis:2b} respectively:
\begin{align}
   p_{+j}  &=  \underset{i}{\max}(q'_{ij}) \label{topsis:2a}\\
   p_{-j}  &= \underset{i}{\min}(q'_{ij})  \label{topsis:2b}
\end{align}

\item Compute the distances to the positive ideal  solution and ($s_{i+}$) and the negative ideal  solution ($s_{i-}$). The distance of each option \textbf{$q'$} to the ideal solution \textbf{$p_{+j}$} and the ideal negative solution \textbf{$p_{-j}$} is given by  Equations \ref{topsis:3a} and \ref{topsis:3b}:
\begin{align} 
    s_{i+} &= \sqrt{\sum_{j=1}^{n} (q'_{ij} - p_{+j})^2} \label{topsis:3a}  \hspace{1cm} and \\
    s_{i-} &= \sqrt{\sum_{j=1}^{n} (q'_{ij} - p_{-j})^2} \label{topsis:3b}
\end{align}

\item Calculate the relative closeness to the ideal solution. The relative closeness  of $\textbf{q}$ to \textbf{$p_{+j}$} and  \textbf{$p_{+j}$} represented by ($c_{i+}$) can be calculated according to Equation \ref{topsis:4}.
\begin{align}
    c_{i+} &= \dfrac{s_{i-}}{s_{i+} - s{i-}} \label{topsis:4}
\end{align}

\item Sort options \textbf{$q_{i}$} in increasing order according to the relative closeness to $c_{i+}$.  
\end{enumerate}

\vspace{-0.5cm}

%TODO: Reescrever o final deste algoritmo de acordo com o padrão passo 5
\subsection{VIKOR} \label{sec:VIKOR}

 The basic concepts of VIKOR is a compromise programming used to get the most satisfactory option by the results of the individual and group regrets. This method has been widely used in several applications fields, such as: location selection, environmental policy and data envelopment analysis \cite{Huang:2009}. 

 %The VlseKriterijumska Optimizacija I Kompromisno Resenje (VIKOR) 
 VIKOR introduces the multicriteria ranking index based on the particular measure of “closeness” to the “ideal” solution. The alternatives are evaluated according to all established criteria  and ranks them according to: i) the minimal distance to the ideal point, ii) the maximum group utility for the “majority” and, iii) the minimum individual regret of the opponent. VIKOR algorithm can be summarized according to the follow steps \cite{Tzeng:2011}, \cite{ Opricovic:2004}:
 
 \begin{enumerate}
 \item Determine the best and the worst  values for all criteria in  $\textbf{Q}$. For a maximization criterion,  the best and worst criteria values represented by $q^{*}_{j}$ and $q^{-}_{j}$ can be calculated respectively according to Equations \ref{vikor:1a} and \ref{vikor:1b}:
 
 \begin{align}
   q^{*}_{j}  &=  \underset{i}{\max}(q'_{ij}) \label{vikor:1a}\\
   q^{-}_{j}  &= \underset{i}{\min}(q'_{ij})  \label{vikor:1b}
 \end{align}
 
 \item Compute the utility measure and the regret measure. The utility measure represented by $S_{i}$ is used to show the average gap of our options and can be calculated according to Equation~\ref{vikor:2a}, where $w_{j}$ corresponds to the  criteria weights, expressing their relative importance. A regret measure represented by $R_{i}$ is used to show the maximal gap for improvement priority and it can be calculated according to Equation \ref{vikor:2b}.
 
 \begin{align}
  S_{i} &= \sum_{j=1}^{n} \dfrac {w_{j} |q^{*}_{j} - q_{ij}|}{|q^{*}_{j} - q^{-}_{j}|} \label{vikor:2a}\\
   R_{i} &= \underset{j}{\max} \dfrac {w_{j} |q^{*}_{j} - q_{ij}|}{|q^{*}_{j} - q^{-}_{j}|}\label{vikor:2b}
 \end{align}

% \item Compute the values $S_{i}$ to compute the average objective gap and $R_{i}$ to compute the maximal gap
% \begin{equation}
% $S_{i} = \sum_{j=1}^{n} \dfrac {w_{j} \cdot |q^{*}_{j} - q_{ij}|}{|q^{*}_{j} - q^{-}_{j}|}$ 
% \end{equation}
% \begin{equation}
% $R_{i} = \underset{j}{\max}$ \dfrac {w_{j} \cdot |q^{*}_{j} - q_{ij}|}{|q^{*}_{j} - q^{-}_{j}|}
% \end{equation}
 
 \item Compute the group utility represented by $Q_{i}$ of each solution. The \textbf{$v$} parameter is used to represent the weight of the strategy of "the majority of criteria". Equation \ref{vikor:3} is used to calculate $Q_{i}$.
 
\begin{align}
    Q_{i} = \dfrac{v \cdot (S_{i} - S^{*})}{(S^{-} - S^{*})} + \dfrac{(1-v) \cdot (R_{i} - R^{*})}{R^{-} - R^{*}}, \label{vikor:3}
\end{align}
 
 where
 
 \begin{align*}
 &S^{*} = \underset{j}{\min} S_{i} \text{ and }
 S^{-} = \underset{j}{\max} S_{i} \\
 &R^{*} = \underset{j}{\min} R_{i} \text{ and } 
 R^{-} = \underset{j}{\max} R_{i} \\
 &v = 0.5
 \end{align*}
 
 %\begin{equation}
 %$Q_{i} = \dfrac{v \cdot (S_{i} - S^{*})}{(S^{-} - S^{*})} + \dfrac{(1-v) \cdot (R_{i} - R^{*})}{R^{-} - R^{*}}$
 %\end{equation}

%where $S^{*} = \underset{j}{\min}$ $S_{i}$,
%$S^{-} = \underset{j}{\max}$ $S_{i}$,
%$R^{*} = \underset{j}{\min}$ $R_{i}$,
%$R^{-} = \underset{j}{\max}$ $R_{i}$ and $v=0.5$ .

 \item Sort options \textbf{$q_{i}$} in decreasing order according to the values $S_{i}$, $R_{i}$ and $Q_{i}$. The results are three ranking lists.
 
 %\item Rank the solutions \textbf{$q$} according to the best measure, \textbf{Q(minimum)}, if the follow conditions are satisfied:
 
 \item Propose as a compromise solution the alternative \textbf{$q$}, which is ranked the best by the measure \textbf{Q(minimum)} if the following two conditions are satisfied:
 
    \textbf{C1.} Acceptable advantage:
    \begin{align*}
        &q_{i + 1}  - q_{i} \ge DQ, 
    \end{align*}
     where $DQ = \dfrac{1}{(N - 1)}$; and N is the number of options
    
    \textbf{C2.} Acceptable stability in decision making: 
    
    The alternative $q_{i}$ must also be the best ranked by S or/and R.
 
    If one of the conditions is not satisfied, then a set of compromise solutions is proposed, which consists of:
    \begin{itemize}
    \item Alternative $q_{i}$ and $q_{i+1}$ if only condition C2 is not satisfied, or
    \item Alternative  $q_{i}$, $q_{i+1}$, ... , $q_{n}$ if condition C1 is not satisfied; and $q^{n}$ is determined by the relation   $q_{n}  - q_{i} < DQ$ for maximum $n$.
    \end{itemize}
 
 \end{enumerate}
 
%This second category is built around methods which utilize various ways to assess the relative importance of multiple attributes and alternatives. Under this category, most the methods were concentrated on weight determination.
 
 In summary, the methods presented in this Section are used to compute the relative importance of multiple criteria and solutions based on an weighting strategy. They have been successfully applied to several real-world scenarios, where multiple conflicting objectives should be satisfied. Next Section describe how the quality of the selection of sensors provided by each algorithm can be evaluated. We also, present all performed experiments and the environment where they were executed.
 
\subsection{Proposal of Evaluation of MCDA methods} \label{sec:evaluation}
\begin{figure}[!htp]
\centering
\includegraphics[scale=0.40]{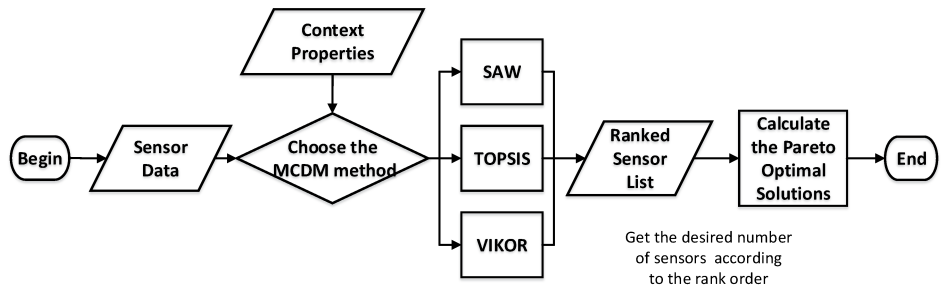}
\caption{Evaluation Workflow}
\label{pic:workflow}
\end{figure}

%\section{Research Methodology} \label{sec:PerformanceEvaluation}
This Section presents the research methodology used in the experiments. As a base of our study we assume the SAW algorithm used by Gao et al. \cite{Gao:2014} and compare it with other popular MCDA (TOPSIS and VIKOR) algorithms.  Our evaluation approach is based on a set of sensor data that will be ranked according to an MCDA method and context properties. The desired number of sensors are retrieved from the top of the ranked list and the \textit{Pareto}-optimal fronts are calculated. Figure \ref{pic:workflow} synthesizes the whole processing proposal for evaluating MCDAs. 

The \textit{Pareto}-optimality criterion  \cite{Marler:2004} is used to compare the quality of the solutions obtained by each method. It uses the dominance concept to determine when a solution is better then other. For example, given two solutions $x$ and $y$, $x$ dominates $y$ ($x \succeq y$) if two conditions are respected:
\begin{enumerate}
    \item The $x$ solution is better than $y$ in at least one objective function;
    \item The $x$ solution is at least equal to $y$ in all objective functions;
\end{enumerate}

 The set of non-dominated solutions is named \textit{Pareto}-optimal set, which represents the set of optimal available solutions for the problem. The \textit{Pareto} fronts is the set of values of the objective functions of the \textit{Pareto}-optimal solutions set. The solutions that are dominated only for the \textit{Pareto}-optimal solutions are located in the second \textit{Pareto} front. The number of \textit{Pareto} fronts that are used in an experiment are directly proportional to the number of non-dominated solution. In this sense, our evaluation process will consider the number of used sensors in the \textit{Pareto}-optimal set and the number of \textit{Pareto} fronts used by each MCDA solution. The \textit{Pareto} fronts are computed through the \textit{fast-non-dominated-sort} algorithm described by Deb~et~al.~(2002)~\cite{Deb:2002}.
 
 We considered two metrics to evaluate the MCDA methods: i) the number of fronts, which indicates the MCDA method with more non-dominated solutions and; ii) the Overall non-dominated vector generation ratio (ONVGR)~\cite{Collette:2004} metric which shows the number of optimal solutions in the Pareto front as a proportion of the number of solutions proposed by the MCDA methods in each front. As closer the ONVGR value is to one better is the solution proposed in that front.

%\subsection{Experiment Environment}

The test environment is composed by one physical machine. Table \ref{tab:environment} describe the hardware and software specification used to perform the experiments.
\vspace{-0.2cm}
% Table generated by Excel2LaTeX from sheet 'Plan1'
\begin{table}[htbp]
  \centering
  \scriptsize
  \caption{Physical Environment}
    \begin{tabular}{cc}
    \toprule
    \textbf{Hardware/Software} & \textbf{}{Specification} \\
    \midrule
    \multicolumn{1}{c}{Processador} & \multicolumn{1}{c}{AMD Processor Vishera 4.2 Ghz} \\
    \multicolumn{1}{c}{Memory} & \multicolumn{1}{c}{32 GB RAM DDR3 Corsair Vegeance} \\
    \multicolumn{1}{c}{Hard Disk} & \multicolumn{1}{c}{HD 2TB Seagate Sata III 7200RPM} \\
    \multicolumn{1}{c}{Operating System} & \multicolumn{1}{c}{Linux Ubuntu Server 14.04 64 Bits LTS} \\
    \multicolumn{1}{c}{Java} & \multicolumn{1}{c}{JDK 1.7} \\
    \multicolumn{1}{c}{Database} & \multicolumn{1}{c}{MongoDB 3.0} \\
    \bottomrule
    \end{tabular}%
  \label{tab:environment}%
\end{table}%
\vspace{-0.3cm}
%\subsection{Experiment Methodology}

The experimental methodology was based on four factors: i) the number of sensors descriptions, ii) the MCDA method, iii) the number of selected sensors and iv) the number of context properties required. In this context, the term context properties will be used to refer to the analysed sensor criteria. Table \ref{tab:exp} shows the used experimental factors and levels, where the combination of the levels of each factor gives a total of 45 experiments.  
\vspace{-0.2cm}
\begin{table}[htbp]
\scriptsize
  \centering
  \caption{Factors and levels used in the experiment}
    \begin{tabular}{rc}
    \toprule
    \multicolumn{1}{c}{\textbf{Factor}} & \textbf{Level} \\
    \midrule
    Number of Sensors Descriptions & 100,000 \\
    MCDA Method & SAW, TOPSIS and VIKOR \\
    Number of Selected Sensors & 1,000 , 5,000 and 10,000 \\
    Number of Context Properties & 2,3,4,5 and 6 \\
    \bottomrule
    \end{tabular}%
  \label{tab:exp}%
\end{table}%
\vspace{-0.3cm}
%Pensar em como justificar
We assume that sensor descriptions such as sensor capabilities and measurements (e.g. frequency and power consumption) are based on the 4027A Series from Bird Technologies\footnote[1]{Bird Technologies  -http://www.birdrf.com/}. Similarly, we assume that context data related to each sensor are retrieved from OpenWeatherMap\footnote[2]{OpenWeatherMap - http://openweathermap.org/} and their current properties values used in this experiment (e.g. battery, price, drift and response time) are assumed to be retrieved by software systems that manage such data and are available to be used.

The criteria and objectives functions used to maximize (max($c_{j}$)) or minimize (min($c_{j}$)) the criteria follow this order: max(battery), min(price), min(drift), max(frequency), min(energy consumption), min(response time). 

\vspace{-0.2cm}

\section{Evaluation Results and Lessons Learned} \label{sec:Results}

\vspace{-0.2cm}

In this Section, we present the gathered data of the performed experiments. In order to make the data visualisation and their meaning easier, we will present the results of each method regarding the number of context properties. %and a result analysis in the end.

\subsection{Evaluation Results}

We will analyse the results regarding the SAW, TOPSIS and VIKOR methods. Two graphics represent the number of used fronts and the ONGVR metric. To represent the number of used front the graphic have two ordinate axis. The abscissa axis has the indexes of the \textit{Pareto} fronts from the first front to the last one .
The left ordinate axis presents the number of solutions retrieved by each method (different colours lines) from each \textit{Pareto} front. %While the y right axis corresponds to the number of optimal solutions available in each \textit{Pareto} front. 
The right ordinate axis corresponds to number of \textit{Pareto} fronts needed to cover a given subset of sensors. To represent the ONVGR metric a graphic with one ordinate axis and one abscissa axis is used. The ordinate axis corresponds to the ONVGR value and the abscissa axis has the indexes of the \textit{Pareto} fronts.

\subsubsection{Selection using six context properties: }\label{sec:6c}

\begin{figure}[htbp]
\center
\includegraphics[scale=0.48]{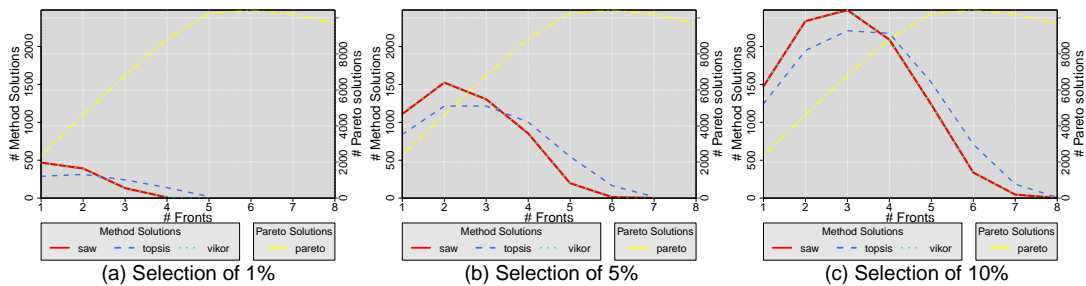}
\caption{\textit{Pareto} fronts for six context properties}
\label{pic:6}
\end{figure}

Figure~\ref{pic:6} presents the quality behaviour of the selection of 1,000 (Figure~\ref{pic:6}.a), 5,000 (Figure~\ref{pic:6}.b) and 10,000 (Figure~\ref{pic:6}.c) of available sensors considering six context properties (as defined in Section~\ref{sec:evaluation}). The number of \textit{Pareto} front slightly increases as the number of selected sensors is raised. Also, the number of optimal sensors available in each front increases according to the number of selected sensors. The MCDA methods concentrates the major part of the solutions in the first fronts due to a high number of conflicts between the used criteria.

\begin{figure}[htbp]
\center
\includegraphics[scale=0.48]{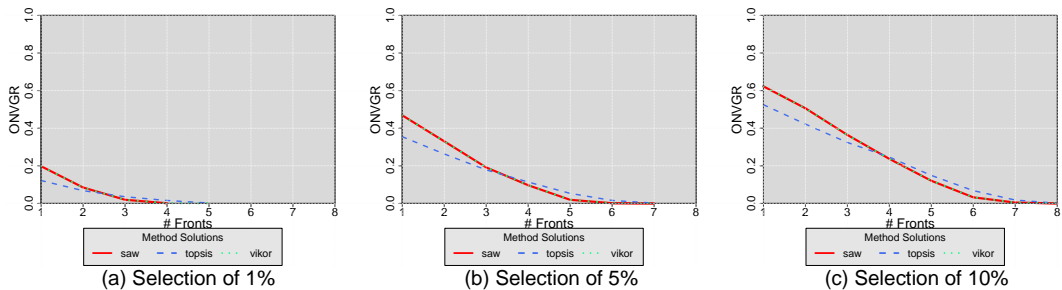}
\caption{ONVGR metric for six context properties}
\label{pic:6-onvgr}
\end{figure}

Figure~\ref{pic:6-onvgr} shows the ONVGR value in each front of the selection of 1,000 (Figure~\ref{pic:6-onvgr}.a), 5,000 (Figure~\ref{pic:6-onvgr}.b) and 10,000 (Figure~\ref{pic:6-onvgr}.c). The ratio value in the first fronts increases proportional to the number of selected sensors. On the other hand the ratio value shows a high loss of optimal sensors, as the ratio values changes from 0.2 to 0.6 in the worst and best scenarios respectively.

The MCDA methods does not use the \textit{Pareto} optimality concept to select the sensors. They aim to select sensors that present a certain level of stability between the context properties values. While, the  \textit{Pareto} optimality solutions do not care about the stability between the context properties values but try to get the greatest number of context properties with the best possible values. 

%In addition, the TOPSIS method presented a worse solution as it has a low ratio value than SAW and VIKOR methods. The SAW method is slightly better than VIKOR method because, although they present practically the same ratio, SAW uses less fronts in some scenarios.

Regarding the MCDA methods, the TOPSIS method presented the worst solution as it shows the lowest ratio value of the analysed MCDA methods in all scenarios. In addition, the SAW method is slightly better than VIKOR method when 1\% of the sensors were desired as it uses less fronts, while the proposed solutions when 5\% and 10\% of the selected sensors were  equivalent.

\subsubsection{Selection using four and five context properties: }\label{sec:4and5}

\begin{figure}[htbp]
\center
\includegraphics[scale=0.48]{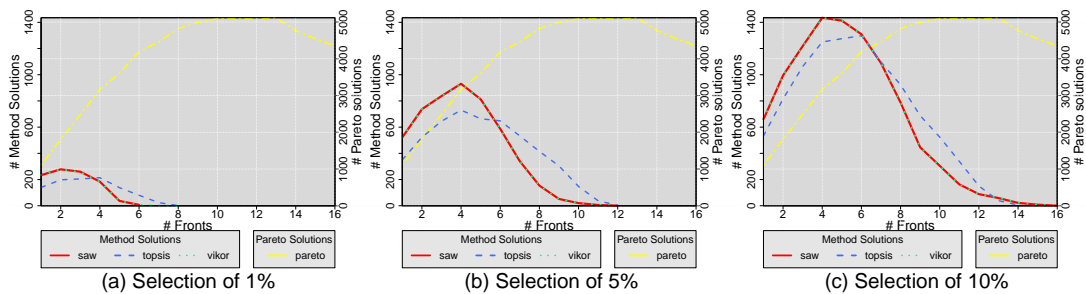}
\caption{\textit{Pareto} fronts for four context properties}
\label{pic:4}
\end{figure}

\begin{figure}[htbp]
\center
\includegraphics[scale=0.48]{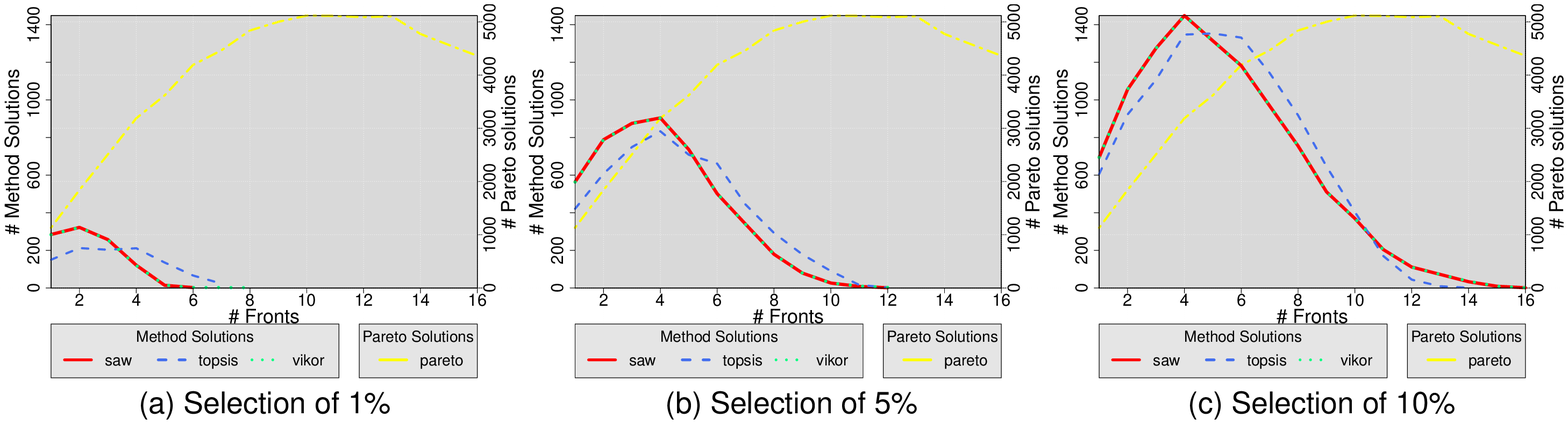}
\caption{\textit{Pareto} fronts for five context properties}
\label{pic:5}
\end{figure}

Figures \ref{pic:4} and \ref{pic:5} presents the quality behaviour of the selection of 1,000 (Figure~\ref{pic:4}.a and~\ref{pic:5}.a), 5,000 (Figure~\ref{pic:4}.b and~\ref{pic:5}.b) and 10,000 (Figure~\ref{pic:4}.c and~\ref{pic:5}.c) of available sensors considering four and five context properties respectively . Analogous to Section~\ref{sec:6c}, the number of \textit{Pareto} front and the number of optimal solutions increases proportional to the number of selected sensors. For four and five context properties the number of \textit{Pareto} fronts is twice as the results presented in Section~\ref{sec:6c} and are not so different, it varies from 6 to 16.

%Figures \ref{pic:4} and \ref{pic:5} presents the quality behaviour of the selection of 1,000 (Figure \ref{pic:4}.a and \ref{pic:5}.a), 5,000(Figure \ref{pic:4}.b and \ref{pic:5}.b) and 10,000 (Figure \ref{pic:4}.c and \ref{pic:5}.c) of available sensors considering four and five context properties respectively. Using four or five context properties, the number of Pareto front increases proportional to the number of selected sensors but the ratio between the number of selected sensors and number of Pareto front decreases even more in relation to \ref{pic:3}. Furthermore, the number of Pareto frontiers establish when we change increase the level of context properties from four to five. 

\begin{figure}[htbp]
\center
\includegraphics[scale=0.48]{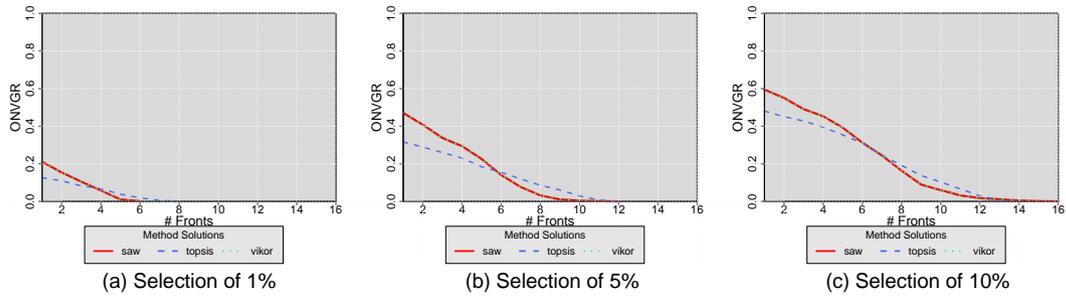}
\caption{ONVGR metric for four context properties}
\label{pic:4-onvgr}
\end{figure}

\begin{figure}[htbp]
\center
\includegraphics[scale=0.48]{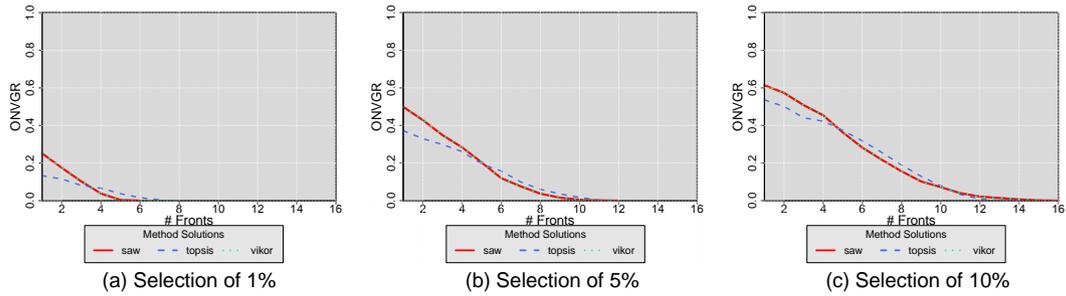}
\caption{ONVGR metric for five context properties}
\label{pic:5-onvgr}
\end{figure}

Figure~\ref{pic:4-onvgr} and~\ref{pic:5-onvgr} show the ONVGR value in each front of the selection of 1,000 (Figure~\ref{pic:4-onvgr}.a and~\ref{pic:5-onvgr}.a), 5,000~(Figure~\ref{pic:4-onvgr}.b and~\ref{pic:5-onvgr}.b) and 10,000 (Figure~\ref{pic:4-onvgr}.c and~\ref{pic:5-onvgr}.c). Although the number of solutions found by each method and the number of \textit{Pareto} solutions are different from Section~\ref{sec:6c}, the ratio between the number of selected sensors and the number of \textit{Pareto} solutions are practically the same as presented in Figure~\ref{pic:6-onvgr}. They also show a low ratio value that changes approximately from 0.2 to 0.6 in the best and worst scenarios respectively.

Considering the MCDA methods, the solution proposed by the SAW method is again slightly better than the solution proposed by the VIKOR method when 1\% of the sensors were desired; while the proposed solutions when 5\% and 10\% of the selected sensors were  equivalent. On the other hand, TOPSIS presents a lower quality solutions as it shows a minor number of sensors in the top first fronts.

\subsubsection{Selection using three context properties: }\label{sec:3}

\begin{figure}[htbp]
	\center
    \includegraphics[scale=0.48]{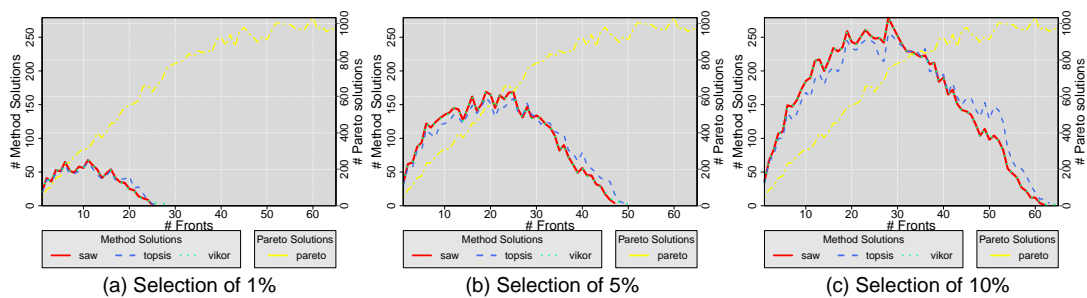}
	\caption{\textit{Pareto} fronts for three context properties}
	\label{pic:3}
\end{figure}

Figure~\ref{pic:3} presents the quality behaviour of the selection of 1,000 (Figure~\ref{pic:3}.a), 5,000(Figure~\ref{pic:3}.b) and 10,000 (Figure~\ref{pic:3}.c) of available sensors considering three context properties. As seen in Section~\ref{sec:4and5}, the number of \textit{Pareto} front increases proportional to the number of selected sensors. This observation are justified because with less context properties we also reduce the number of context properties conflicts, the number of \textit{Pareto} optimal solutions per front and the number of solutions found per front, which increases the probability for finding solutions with a higher level of stability.

\begin{figure}[htbp]
	\center
    \includegraphics[scale=0.48]{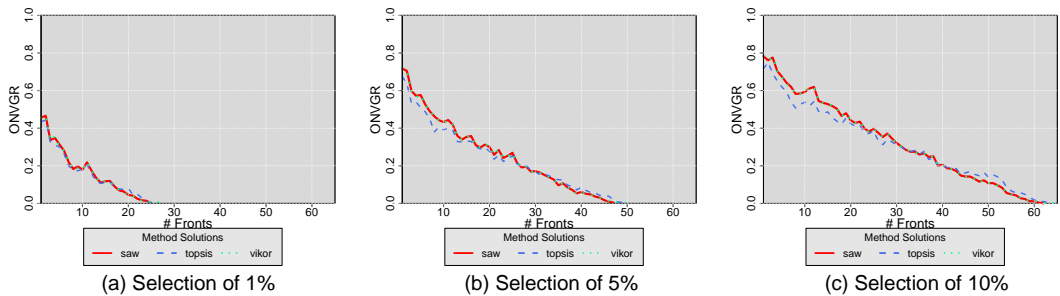}
	\caption{ONVGR metric for  three context properties}
	\label{pic:3-onvgr}
\end{figure}

Figure~\ref{pic:3-onvgr} shows the ONVGR value in each front of the selection of 1,000 (Figure~\ref{pic:3-onvgr}.a), 5,000(Figure~\ref{pic:3-onvgr}.b) and 10,000 (Figure~\ref{pic:3-onvgr}.c). The ratio value in the first fronts are slightly higher than the ratio values presented in Section~\ref{sec:4and5} due to the reduction of the number of criteria. In this sense, the ratio values changes from approximately 0.4 to 0.8 in the worst and best scenarios respectively.

Moreover, when the MCDA methods are analysed the quality of the solution proposed by the SAW method was slighter better than the quality of the solution proposed by VIKOR method when 1\%, 5\% and 10\% of the available sensors were selected as the SAW solution uses less \textit{Pareto} fronts. Similar to Section~\ref{sec:4and5}, the TOPSIS method presented the solution with low quality as it had less solutions than the SAW and VIKOR methods in the top first fronts.

\subsubsection{Selection using two context properties: }

\begin{figure}[htbp]
	\center
    \includegraphics[scale=0.48]{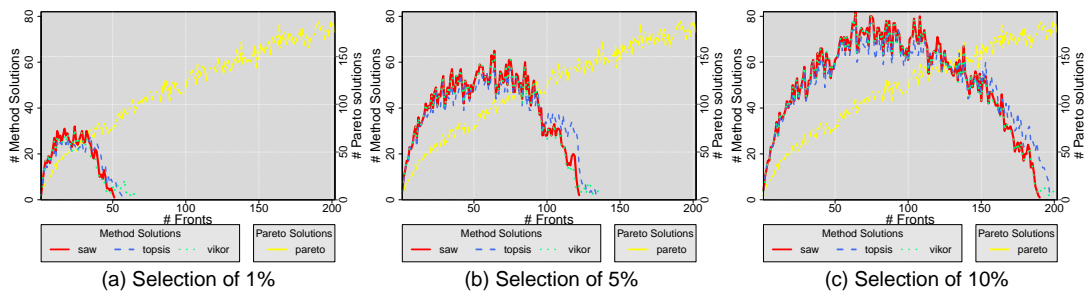}
	\caption{\textit{Pareto} fronts for two context properties}
	\label{pic:2}
\end{figure}

Figure \ref{pic:2} presents the quality behaviour of the selection of 1,000 (Figure~\ref{pic:2}.a), 5,000(Figure~\ref{pic:2}.b) and 10,000 (Figure~\ref{pic:2}.c) of available sensors considering two context properties. Analogous to Section~\ref{pic:3}, the number of \textit{Pareto} fronts and the number of optimal solutions increases directly proportional to the number of selected sensors.
    
\begin{figure}[htbp]
	\center
    \includegraphics[scale=0.48]{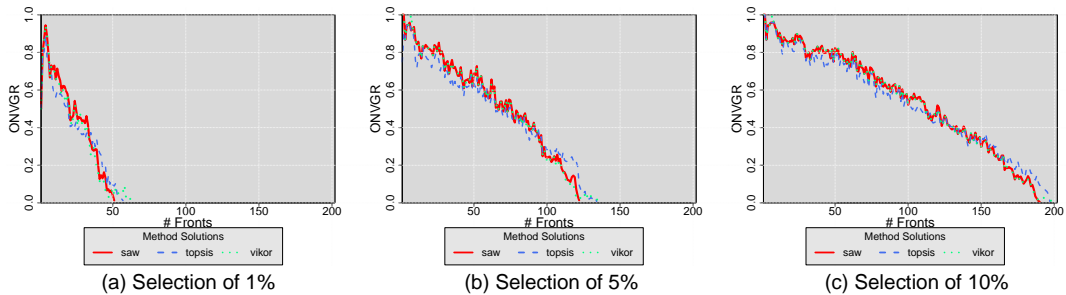}
	\caption{ONVGR metric for  two context properties}
	\label{pic:2-onvgr}
\end{figure}

Figure~\ref{pic:2-onvgr} shows the ONVGR value in each front of the selection of 1,000 (Figure~\ref{pic:2-onvgr}.a), 5,000 (Figure~\ref{pic:2-onvgr}.b) and 10,000 (Figure~\ref{pic:2-onvgr}.c). Due to the minimal number of criteria conflicts, the ratio value tends to get closer to one as it less optimal solutions are available. In this sense, the ratio changes from approximately 0.8 to 1 in the worst and best scenarios respectively, which shows that all optimal solutions are selected.

Furthermore, the SAW method presented again the solution with better quality independently of the selected sensors numbers as it presented a high ONVGR value and uses less fronts than VIKOR and TOPSIS. The solution presented by VIKOR was quite similar to the solution presented by the SAW method, but it solution uses more fronts than SAW. The TOPSIS method presented the poorest solution, as it shows a higher number of fronts and a low ONGVG value in the top first fronts.

%Using two context properties, the number of Pareto front increases proportional to the number of selected sensors. Also, the number of Pareto optimal solutions presented in each front is practically the same solution presented by the analysed methods. In this scenario, all methods were able to propose the best available sensors.

\subsection{Lessons Learned}

%In this section, we have compared the behaviour and quality of different MCDA methods for sensor search and selection.Our result have shown that as the number of context properties increases the number of Pareto front decreases and the number of optimal solutions in each front also increases. Besides, the ratio between the selected sensors and the available sensors in each Pareto front decreases as the number of context properties increases. Thus, as more context properties are used it becomes more difficult to retrieve the optimal solutions. Moreover, as we raise the number of selected sensors for a set of context properties, we increase the number of optimal solutions in each Pareto front and the number of Pareto fronts. 

%The methods have shown practically the same behaviour when two context properties were used. On the other hand, when three or more context properties were used the SAW and VIKOR methods selected the major number of sensors in the first fronts while the TOPSIS method selected a major number of sensors in the last front. Moreover, as the number of selected sensors increases for a set of context properties the behaviour of the analysed methods tend to look similar.

In this section, we have compared the behaviour and quality of different MCDA methods for sensor search and selection. Firstly, it is important to highlight the number of optimal solutions available in each scenario. As expected, the number of optimal solutions  increases proportional to the number of fronts. It occurs due to the non-dominated solution concept used to compute the optimal solutions set in each front. In this sense, the number of optimal solutions is not influenced by the number of selected sensors.

On the other hand, the number of selected sensor affects the number of optimal solutions that are founded by the MCDA algorithms. The influence of the number of selected sensors can be justified, because it increases the chances of the MCDA find the optimal sensors set. In all scenarios, the ONVGR metric clearly shows the significant increase of the ratio between the number of optimal sensors and the number of sensors found by each MCDA algorithm when more sensors are selected. 

%Não está legal é necessário melhorar
The context properties also influence the number of optimal solutions obtained by each MCDA algorithm. The number of context properties is directly proportional to number of optimal solutions available in each front. It is because as more context properties are used the number of conflicts between the criteria increases and consequently the number of non-dominated solutions increases. In other words, we reduce the chances to find a small set of solutions which present the best trade-off between the analysed context-properties.

Also, the ONVGR metric allows to compare how the number of context properties influence the number of optimal selected sensors. Although the number of selected sensors in each front is different for six, five or four context properties the ONVGR value is practically the same for all and indicates that a low number of optimal sensors is founded in each one. When three or two context properties are used, the ONVGR value is higher for all scenarios and consequently a higher number of optimal sensors is founded when less context properties is used.
 
%Regarding the analysed MCDA methods it is possible to observe that for all analysed scenarios the SAW method presented at least an equal number of fronts and the ONVGR value  than TOPSIS and VIKOR method. The VIKOR method presented a solution pretty closer to the proposed solution by SAW algorithm but in some scenarios it solution has more fronts. Finally, the TOPSIS method presented the poorest solution as in the major part of the scenarios it uses more fronts and presented a low ONVGR value than SAW and VIKOR methods.

Regarding the analysed MCDA methods it is possible to observe that for all analysed scenarios the SAW method, which uses regular arithmetical operations of multiplication and addition to rank the options, presented at least an equal number of fronts and the ONVGR value than TOPSIS and VIKOR method. The VIKOR method, which apply the compromise programming concept providing a maximum ‘‘group utility’’ for the ‘‘majority’’ and a minimum of an individual regret for the ‘‘opponent’’, presented a solution pretty closer to the proposed solution by SAW algorithm but in some scenarios it solution has more fronts. Finally, the TOPSIS method which ranks the solutions according to the distance to the ideal solution and the greatest distance from the negative-ideal solution without consider the relative importance of these distances,  presented the poorest solution as in the major part of the scenarios it uses more fronts and presented a low ONVGR value than SAW and VIKOR methods.

%\subsection{Result Analysis}

%Our result analysis have showed that as the number of context properties increases the number of Pareto front decreases and the number of optimal solutions in each front also increases. Besides, the ratio between the selected sensors and the available sensors in each Pareto front decreases as the number of context properties increases. Thus, as more context properties are used more difficult is to retrieve the optimal solutions. Also, as we raise the number of selected sensors for a set of context properties we increase the number of optimal solutions in each Pareto front and the number of Pareto fronts. 

%Finally, the methods have showed practically the same behaviour when two context properties were used. On the other hand, when three or more context properties are used the SAW and VIKOR methods selected the major number of sensors in the first fronts while the TOPSIS method selected a major number of sensors in the last front. Moreover, as the number of selected sensors increases for a set of context properties the behaviour of the analysed methods they tend to look similar.

\vspace{-0.4cm}

\section{Related Work} \label{sec:RelatedWork}

\vspace{-0.2cm}

Today there are several approaches that enable the sensor management. Perera et al. \cite{Perera:2014a} and R\"{o}mer et al. \cite{Romer:2010} present surveys that describes several techniques, methods, models, features, systems, applications, and middleware solutions related to the IoT context. These surveys shows that the algorithms used to perform the sensor search and selection can be splitted in two groups: prediction models and keyword or context information. In this Section we present the main work related to each group. 

Elahi et al. \cite{Elahi:2009} presents a primitive called \textit{sensor ranking} to perform the sensor search in an efficient way. The main idea of \textit{sensor ranking} primitive is to explore the periodicity presented by the sensor in some cases using prediction models that rank the sensors according to the probability to meet a user query. The Single-Period and the Multi-Period predictions models are used in this paper and the gathered data allow to observe a performance improvement to select the sensors.
Ostermaier et al. \cite{Ostermaier:2010} present a search engine for the Web of Things called \textit{Dyser} to conduct searches in scalable environments with highly dynamic content. \textit{Dyser} is able to collect and store data and information from sensors to allow search based on metadata. It also extends the work presented by Elahi et. al. \cite{Elahi:2009} using the Aggregated Prediction Model. The results showed that the algorithms presented a better quality selection when compared with the random model.

Truong et al. \cite{Truong:2013} also extends the work presented by Elahi et. al. \cite{Elahi:2009}  and propose a prediction model based on fuzzy logic named  Time-Independent Prediction Model. This model is able to detect anomalies about sensor behavior using metrics of density and stability. The density metric is used to estimate the  probability of a certain value belong to a specific sensor while the stability metric estimates the stability of these sensors in the past. The combination of these metrics allow to rank the sensors and check their state. Thus, the solution presented is able to reduce the necessary communication for sensor search and selection.

Carlson and Schrader \cite{Carlson:2014} present a search engine named Ambient Ocean to search and select sensors using context information. The search engine uses metadata, which is stored in a global repository, to establish the sensors context and carry out the search in a more efficient and effective manner. Ambient Ocean uses multi-task similarity models based on the Weighted Slope One algorithm to select the sensors.  In scenarios where the characteristics of the sensors are difficult to model, collaborative filtering techniques are employed to compute similarities between users or sensors based on information history.

Ding et al. \cite{Ding:2012} propose a hybrid search engine to IoT environments, able to perform searches using quantitative values, keywords and spatio-temporal relations. The architecture of this search engine is based on a bottom-up model with three layers, the first layer is responsible for sensing and monitor the equipment. The second layer is responsible to store the data in a distributed form. The third layer provides optimized access to data from the sensors. The search for keywords and quantitative values is optimized by a B+ tree and the search base on time-space relationships uses a R tree. This search engine allows the discovery of the objects state at run-time as the sensors sends continuous data to the storage layer, which index these data according to data-structure used.

 Guinard et al. \cite{Guinard:2010} propose a module for the integration architecture named SOCRADES, which aims to enable ubiquitous integration services running on embedded with other business processes devices. The proposed module is based on the model Publish/Subscribe and uses a global repository to store meta-data about the available devices. The repository works with a monitor that is responsible to update the devices states and their QoS attributes. The sensor search is made by keywords and is sorted according to the QoS attributes prioritized by the user.
 Kothari et al. \cite{Kothari:2014} presents an architecture denominated  DQS-Cloud to optimize the sensor search, provide resilience to faults and QoS degradation and also optimize system performance managing sensor data streams. The sensor search is based on keywords and considers the QoS attributes specified by users. Moreover, in order to reduce communication overhead, the authors proposes an optimization mechanism to reuse sensors flows to similar requests. The results showed that the optimization module is able to reduce the bandwidth and processing rate of the providers.
 
 Shah et al. \cite{Shah:2012} presents a search mechanism based on Coordinate Virtual System to find process in P2P networks. A coordinate is assigned to a node representing a physical location in relation to other nodes. The sensor search uses keywords and the returned sensors are ranked according to the euclidean distance to the QoS attributes specified by the user. A qualitative approach shows that the proposed search mechanism was the only one able to perform a precision query at real time.
 Ruta et al. \cite{Ruta:2013} proposes a framework to manage semantic notations of data streams, devices, high level events and services. The requests uses the CoAP protocol based on the RESTful architectural style, which allow to use inference to support the sensor search and their compositions. A data mining mechanism was used to retrieve the sensor search in real time to improve the sensor selection. The sensor selection is based in the Concept Covering inference followed by a ranking algorithm.
 
 Perera et al. \cite{Perera:2013} proposes a framework named CASSARAM, which performs the sensor search and selection regarding the QoS attributes specified by the user. The selection process is divided in two phases. In the first phase, the static sensor attributes, such as manufacturer or type, are used to limit the user space search. In the second phase, the result query of the first phase is evaluated in a multi-dimensional space where each axis corresponds to a QoS user attribute. The sensors are indexed regarding the Comparative Priority-based Weighted Index and ranked according to their euclidean distance to the optimal point. The authors also proposes a heuristic named Comparative Priority-based Heuristic Filtering, which removes the sensors that are far from the ideal point prioritizing the TOP-K selection. The results shows that using up to 10,000 sensors, the framework presents a satisfactory performance with a high precision.
 
 Gao et al. \cite{Gao:2014}  proposes the Automated Complex Event Implementation System to manages  different run-time data streams. The sensors and their data streams are described according to SSN ontology and stored in a repository with their QoS attributes. The system acts such as a middleware between a sensor data stream and an application. The middleware are able to perform the sensor search and select using the Simple-Additive-Weighting algorithm to find the best trade-off between the specified QoS attributes. 
 
 \vspace{-0.6cm}
 
 \begin{table}[htbp]
  \centering
  \scriptsize
  \caption{Related works summary}
    \begin{tabular}{ccccc}
    % \hline
    \toprule
    \multicolumn{1}{c}{\textbf{Paper}} & \textbf{Search Technique} & \textbf{Search} & \textbf{Selection Method} & \textbf{QoS} \\
    \midrule
    % \hline
    \multirow{2}[2]{*}{\cite{Elahi:2009}} & \multirow{2}[2]{*}{Prediction Model} & \multirow{2}[2]{*}{Sensor State} & Single-Period and Multi-Period & \multirow{2}[2]{*}{No} \\
          &       &       &   Prediction model &  \\
    % \hline
    \cite{Ostermaier:2010} & Prediction Model & Sensor State & Aggregated Prediction model  & No \\
    % \hline
    %\cite{Pfisterer:2011} & Prediction Model & Sensor State & Not specified & No \\
    % \hline
    \cite{Truong:2013} & Prediction Model & Sensor State & Time-Independent Prediction Model & No \\
    % \hline
    \cite{Carlson:2014} & Prediction Model & Context Information & Weighted Slope One Algorithm & No \\
    % \hline
    \multirow{2}[2]{*}{\cite{Ding:2012}} & \multirow{2}[2]{*}{Index} & Keywords, Sensor State & B+ tree and  & \multirow{2}[2]{*}{No} \\
          &       & Context Information & R tree &  \\
    % \hline
    \cite{Guinard:2010} & Score and Ranking & Keywords & Not specified & Yes \\
    % \hline
    \multirow{2}[2]{*}{\cite{Kothari:2014}} & Score, Ranking & \multirow{2}[2]{*}{Keywords} & \multirow{2}[2]{*}{Not specified} & \multirow{2}[2]{*}{Yes} \\
          &  and Similarity &       &       &  \\
    % \hline
    \cite{Ruta:2013} & Inference and Ranking & Keywords & Concept Covering & No \\
    % \hline
    \cite{Shah:2012} & Score and Ranking & Keywords & Euclidean Distance & Yes \\
    % \hline
    \multirow{2}[2]{*}{\cite{Perera:2013}} & \multirow{2}[2]{*}{Score and Ranking} & \multirow{2}[2]{*}{Context Information} & Euclidean Distance and Comparative & \multirow{2}[2]{*}{Yes} \\
          &       &       &  Priority-based Heuristic Filtering &  \\
    % \hline
    \cite{Gao:2014}  & Score and Ranking & Context Information & Simple-Additive-Weighting & Yes \\
    % \hline
    \bottomrule
    \end{tabular}%
  \label{tab:RelatedWork}%
\end{table}%

\vspace{-0.3cm}

The Table \ref{tab:RelatedWork} summarizes the main characteristics of the works presented in this Section. \cite{Elahi:2009}, \cite{Ostermaier:2010}, \cite{Truong:2013} e \cite{Carlson:2014} uses prediction models and do not consider QoS attributes to choose the sensors as they are just interesting in the sensor state.  \cite{Ding:2012} also do not consider QoS attributes but worry about to offer efficient data structures to store the sensor state at run-time. On the other hand, \cite{Guinard:2010} e \cite{Kothari:2014} highlight the importance to select the sensors based on their QoS properties, but do not present a specific method for sensor search and selection. \cite{Shah:2012},  \cite{Perera:2013} and \cite{Gao:2014} uses methods to score and ranking their sensors. \cite{Shah:2012} and \cite{Perera:2013} use the Euclidean distance of the sensor to the optimal point to score and rank while \cite{Gao:2014} applies the SAW method. 

Briefly, these works present different mechanisms to perform the sensor search and selection considering QoS properties. However, they do not evaluate and compare the quality of the proposed solutions of a specific technique according to the number of desired QoS properties. Thus, to fulfill this gap we have proposed a methodology to enable the comparison of different sensor search techniques from the quality of search perspective. In addition, a case study considering three sensor search techniques is presented to demonstrate our methodology.

\vspace{-0.2cm}

\section{Conclusion} \label{sec:Conclusion}
The integration of IoTs with cloud computing composes the CoT and poses several new challenges. One of these challenges focus on the sensor search and selection field regarding a set of context properties explicit desired by an user. In this paper, we examined some multi objective decision analysis methods applied in different scenarios. Specifically, we have analysed the \textit{Simple Additive Weight method} (SAW), the \textit{Technique for the Order of Prioritisation by Similarity to Ideal Solution} (TOPSIS) and \textit{VIseKriterijumska Optimizacija I Kompromisno Resenje} (VIKOR) according to the number of desired solutions and the number of context properties. 

We described a methodology that uses the \textit{Pareto} optimality concept to enable comparisons between the solutions proposed by these methods. %The gathered results have shown that the quality of the proposed solutions according to the \textit{Pareto} optimality concept tends to be close when a small number of properties is used and/or the number of desired sensors increases.
The gathered results allowed to observe the impacts of \textit{`the number of context properties'} and  \textit{`the number of desired sensors'} towards the quality of the final solution based on the Pareto optimality concept.

Using the overall non-dominated vector generation ratio (ONVGR) we observed when six, five and four context-properties are used,  the proportion of optimal sensors retrieved is low because of the high number of context-properties conflicts, and also is practically the same for all. In contrast, when three or two context-properties are used the proportion of optimal sensors retrieved increases significantly due to the small number of context-properties conflicts.

In addition, the ONVGR value is directly proportional to the number of selected sensors. In other words as more sensors are selected more optimal sensors are used. Regarding the analysed MCDA methods, we could observe that the SAW presented the solution with better quality as their ONVGR value and number of fronts is equal or better than the values retrieved by TOPSIS and VIKOR. For future work, we will apply these methods in existent centralized and decentralized architectures for IoT where their performance will be measured. Also, new performance metrics and approaches for sensor selection will be analysed for each kind of architecture.

% We have used the \textit{Pareto} optimality concept to compare the quality of the solutions proposed by these methods. The gathered results have shown that the quality of the proposed solutions according to the \textit{Pareto} optimality concept tends to be close as a short number of properties are used and/or the number of desired sensors increase. We also observe when six, five and four context-properties are used the proportion of optimal sensors retrieved in the first front is the same and indicates a low value. In contrast, when three or two context-properties the proportion of optimal sensors retrieved increases significantly.
% For future work, we will apply these methods in existent centralized and decentralized architectures for IoT where their performance will be measured. Also, new performance metrics and approaches for sensor selection will be developed for each kind of architecture.

\vspace{-0.5cm}

\bibliographystyle{wileyj}
\bibliography{bibliograph}

%\section*{I. Appendix}

\end{document}